\documentclass[aps]{revtex4}

\usepackage{enumerate}
\usepackage{enumitem}
\usepackage{graphicx}
\usepackage{dcolumn}
\usepackage{bm}
\usepackage{graphicx}
\usepackage{amssymb}
\usepackage{epstopdf}
\usepackage{xcolor}

\usepackage[english]{babel}
\usepackage{xspace}
\newcommand{\LTO}{$\mathrm{LaTiO}_3$\xspace}
\newcommand{\STO}{$\mathrm{SrTiO}_3$\xspace}
\newcommand{\LAO}{$\mathrm{LaAlO}_3$\xspace}

\newcommand{\LXO}{$\mathrm{LaXO}_3\mathrm{(X=Al,Ti)}$\xspace}
\usepackage{soul}

\begin{document}
\title{\Large Single-band to two-band superconductivity  transition in two-dimensional oxide interfaces. }
\author{G. Singh$^{1,2}$, A. Jouan$^{1,2}$, G. Herranz$^{3}$, M. Scigaj$^{3}$, F. S\'anchez$^{3}$, L. Benfatto$^{4,5}$, S. Caprara$^{5,4}$, M. Grilli$^{5,4}$, G. Saiz$^{1,2}$, F. Couedo$^{1,2}$, C. Feuillet-Palma$^{1,2}$, J. Lesueur$^{1,2}$, N. Bergeal$^{1,2*}$}

\affiliation{$^1$Laboratoire de Physique et d'Etude des Mat\'eriaux, ESPCI Paris, PSL Research University, CNRS, 10 Rue Vauquelin - 75005 Paris, France.\\
$^2$Universit\'e Pierre and Marie Curie, Sorbonne-Universit\'es,75005 Paris, France.\\
$^3$Institut de Ci\'encia de Materials de Barcelona (ICMAB-CSIC), Campus de la UAB, 08193 Bellaterra, Catalonia, Spain.
$^4$Institute for Complex Systems (ISC-CNR),  UOS Sapienza, Piazzale A. Moro 5, 00185 Roma, Italy\\
$^5$Dipartimento di Fisica Universit\`{a} di Roma``La Sapienza'', Piazzale A. Moro 5, I-00185 Roma, Italy.\\}

\maketitle

\indent
\large

\textbf{In multiorbital materials, superconductivity can exhibit new exotic forms that include several coupled condensates. In this context, quantum confinement in two-dimensional superconducting oxide interfaces offers new degrees of freedom to engineer the band structure and selectively control $3d$-orbitals occupancy by electrostatic doping. However, the presence of multiple superconducting condensates in these systems has not yet been demonstrated. Here, we use resonant microwave transport to extract the superfluid stiffness of the (110)-oriented \LAO/\STO interface in the entire phase diagram.  We evidence a transition from single-band to two-band superconductivity driven by electrostatic doping,  which we relate to the filling of the different $3d$-orbitals based on numerical simulations of the quantum well. Interestingly, the superconducting transition temperature decreases while the second band is populated, which challenges the Bardeen-Cooper-Schrieffer theory. To explain this behaviour, we propose that the superconducting order parameters associated with the two bands have opposite signs with respect to each other.} \\

\maketitle

In cubic perovskites, transition metal ions are surrounded by six oxygen ions in an octahedron configuration. Because of the negative charge of O$^{2-}$ ions, the transition metal $3d$ electrons are subject to an anisotropic crystal field, which splits the five $3d$ orbitals into three degenerated $t_{2g}$ orbitals ($d_{xy}$, $d_{xz}$ and $d_{yz}$)  and two degenerated $e_g$ orbitals ($d_{x^2-y^2}$ and $d_{z^2}$) at higher energy \cite{imada}. This situation is encountered in bulk \STO crystals whose conduction band is formed by the coupling between $t_{2g}$ orbitals  at neighbouring Ti lattice sites through the $2p$ orbitals of the oxygen atoms. Meanwhile, the physical properties of two-dimensional \STO-based interfaces such as \LAO/\STO \cite{Ohtomo:2004p442,Caviglia:2008p116} or \LTO/\STO \cite{Biscaras:2010p7764} interfaces are also deeply affected by the further splitting of the $t_{2g}$ bands under quantum confinement. Many theoretical works have shown that a complex band structure involving bands with different orbital symmetries is generated at the interface \cite{popovic,delugas,pavlenko,pentcheva, scopigno}. These predictions are supported by experiments such as x-ray absorption spectroscopy \cite{salluzzo}, optical conductivity \cite{seo},  Hall effect \cite{Kim:2010p9791,Ohtsuka:2010p9619,biscaras2} and quantum oscillations  \cite{Ben ShalomSdH,cavigliaSdH, yang}, which evidence multiband transport at high carrier densities. The presence of several disconnected Fermi surface sheets makes these materials more prone to exhibit unconventional properties. \\
\indent One of the main challenges in understanding superconductivity in multiband systems is to identify the connection between superconductivity and the different band occupancies.  In the conventional weak-coupling BCS theory, the superconducting critical temperature is expressed as $T_c\simeq\hbar\omega_De^{-\frac{1}{\lambda}}$, where  $\hbar\omega_D$ is an energy cutoff (Debye energy in  the standard phonon-mediated pairing mechanism) and $\lambda=N(0)V_0$ is the coupling constant \cite{BCS}. Because of the exponential factor, the density of states  at the Fermi level, $N(0)$ and the pairing potential, $V_0$, are the most relevant parameters in the determination of $T_c$. Although the origin of superconductivity in \STO is still debated, it is widely believed that its gigantic low-temperature dielectric constant $\epsilon_R\simeq25000$ \cite{NEVILLE:1972p3397} should play a pivotal role in the pairing mechanism and consequently in the determination of $V_0$. On the other hand, in a two-dimensional system, the density of states is directly proportional to the effective mass $m^*$of the carriers,  $N(0)=\frac{m^*}{\pi\hbar^2}$. Whereas these two parameters are fixed in a bulk material, oxide interfaces offer several possibilities to vary $N(0)$ and $V_0$, for instance by selecting $t_{2g}$ orbitals through crystal orientation \cite{gervasi} or by electrostatic control of the band fillings.  Here, we use resonant microwave transport to investigate  the superfluid stiffness of the (110)-oriented interface. In contrast with the reported single band superconductivity in the conventional (001)-oriented interfaces \cite{richter, singh,bert2}, we demonstrate the presence of two superconducting coupled condensates at high carrier doping, which we relate to the degenerated $d_{xz}$/$d_{yz}$ bands on the one hand and the $d_{xy}$ band on the other hand. Because of the unusual decrease of $T_c$ concomitant with the emergence of the second band, we propose that the two condensate are coupled by a repulsive interaction leading to opposite-sign gaps s$\pm$-wave superconductivity. \\

 \begin{figure*}[t]
\begin{center}
\vskip 0.5cm
\abovecaptionskip 5cm
\includegraphics [width=16cm]{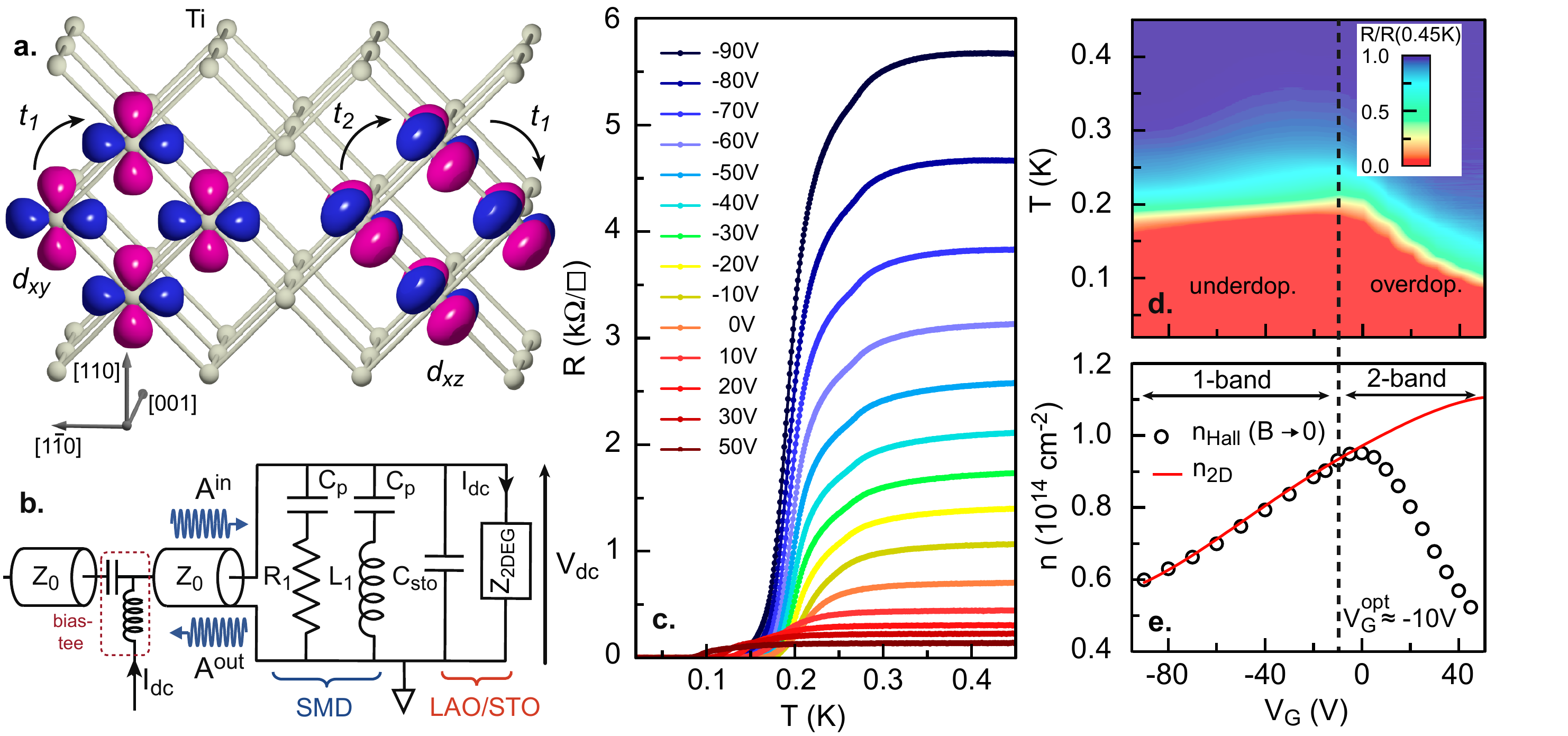}%
\end{center}
\vskip -0.5cm
\caption{\textbf{Superconductivity and multiband transport in (110)-oriented \LAO/\STO interfaces}. a) Scheme of the $t_{2g}$ orbitals in the (110)-oriented \STO. $d_{yz}$ orbitals (not shown) are obtained by a 90$^{\circ}$ rotation of the $d_{xz}$ ones along the (001) direction. $t_1$ and $t_2$ represent the hoping terms along the different directions. In this experiment current was injected along the (1-10) direction. b) Resonant sample circuit comprising the \LAO/\STO heterostructure ($Z_\mathrm{2DEG}$ and $C_\mathrm{sto}$) and the SMD components, $R_1$=100 $\Omega$ and $L_1$= 9 $nH$. $C_p$= 2 $\mu F$ are protective capacitances that avoid dc current to flow through $L_1$ and $R_1$ without influencing the resonance frequency of the circuit $\omega_0$ \cite{singh}. A bias-tee is used to separate the dc signal from the RF one. c) Sheet resistance as a function of temperature measured in dc for different gate voltages in the range [-90V,+50]. (d) Sheet resistance normalized by its value at T=0.45K in color scale as a function of gate voltage and temperature. (e) Hall carrier density $n_\mathrm{Hall}$ measured at $T$=3K in the limit  B$\rightarrow$0 as a function of $V_\mathrm{G}$. The gate dependence of the 2D carrier density $n_\mathrm{2D}$ has been obtained by integrating the gate
capacitance and by matching it to $n_\mathrm{Hall}$  in the one-band regime ($V_\mathrm{G}<-10V$) (see Methods).}
\end{figure*}

While most studies have focused on (001)-oriented \LAO/\STO heterostructures, it has been recently shown that a superconducting two-dimensional electron gas (2-DEG) can also develop in other crystal orientations, such as the (110) direction \cite{gervasi} and the (111) direction \cite{monteiro,rout,davis}. In this study, we used 10 uc-thick \LAO epitaxial layers  grown on 3$\times$3~mm$^2$ (110)-oriented \STO substrate by Pulsed Laser Deposition (see Methods section) \cite{gervasi}. The orientation of the  $t_{2g}$ orbitals in this heterostructure is shown in Fig. 1a. Because of the symmetry of the orbitals, the ability of electrons to hop between two neighbouring Ti lattice sites strongly depends on the spatial directions, which causes anisotropic band properties.  After the growth, a weakly conducting metallic back-gate of resistance $\sim$100 k$\Omega$ was deposited on the backside of the 200 $\mu m$-thick  substrate. The \LAO/\STO heterostructure was inserted in a  microwave circuit board between the central strip of a coplanar waveguide transmission line and its ground to perform microwave measurements as described in reference \cite{singh}.  The Surface Mounted microwave Devices (SMD), which include resistor $R_1$ and inductor $L_1$, were added to form a parallel RLC resonant circuit, where the capacitance $C_\mathrm{STO}$ is the intrinsic capacitance of the \STO substrate (Fig. 1b).
 \indent After cooling the sample to 450 mK, the first positive polarization was applied  to a maximum gate voltage $V_\mathrm{G}$=+50 V  to ensure that no hysteresis would occur upon further gating \cite{biscaras3}.  Figure 1c displays the temperature-dependent sheet resistance $R$ of the 2-DEG,  which was measured in dc for different gate voltages in the range [-90V,+50V].   A clear superconducting transition to a zero resistive state is observed for all gate values. The resulting phase diagram, obtained by plotting  the normalized resistance  in colour scale as a function of temperature and gate voltage is shown in Fig. 1d.  In contrast with the conventional (001)-orientation \cite{Caviglia:2008p116,biscaras2, hurand}, superconductivity can not be suppressed by carrier depletion. In the underdoped (UD) regime, the transition temperature weakly increases with $V_\mathrm{G}$  to a maximum value of $\simeq$ 200 mK at the optimal doping $V_\mathrm{G}^\mathrm{opt}\simeq$-10 V, before  decreasing in the overdoped (OD) regime ($V_\mathrm{G}>V_\mathrm{G}^\mathrm{opt}$).\\

 To understand this unexpected gate dependence, we analyse the Hall effect in the normal state of the 2-DEG. Whereas the Hall resistance $R_\mathrm{H}$ is linear with the magnetic field $B$ at low-doping, corresponding to one-band transport (labelled band 1), this is not the case at high doping (Supplementary Figure 1). This suggests that a one-band to two-band transition occurs in the Hall effect for a certain gate voltage.  The variation of the 2-DEG carrier density $n_\mathrm{2D}$ over the entire phase diagram, can be retrieved by plotting the charging curve of the capacitor (Fig. 1e) (see Methods). The new band (band 2) filling threshold corresponds to the gate value where  $n_\mathrm{Hall}=\frac{B}{eR_\mathrm{H}}$ measured in the limit B$\rightarrow$0, deviates from $n_\mathrm{2D}$ \cite{biscaras2,singhCR}. Interestingly, this change  in Hall behaviour occurs  at the optimal doping point ($V_\mathrm{G}^\mathrm{opt}\simeq-10V$), which establishes a correlation between the OD superconducting regime and the two-band transport regime observed in the normal state.  However, the significant decrease of transition temperature in this regime also indicates that the filling of band 2 is detrimental for superconductivity. \\

  \begin{figure*}[t]
\begin{center}
\vskip 0.8cm
\abovecaptionskip 5cm
\includegraphics [width=13cm]{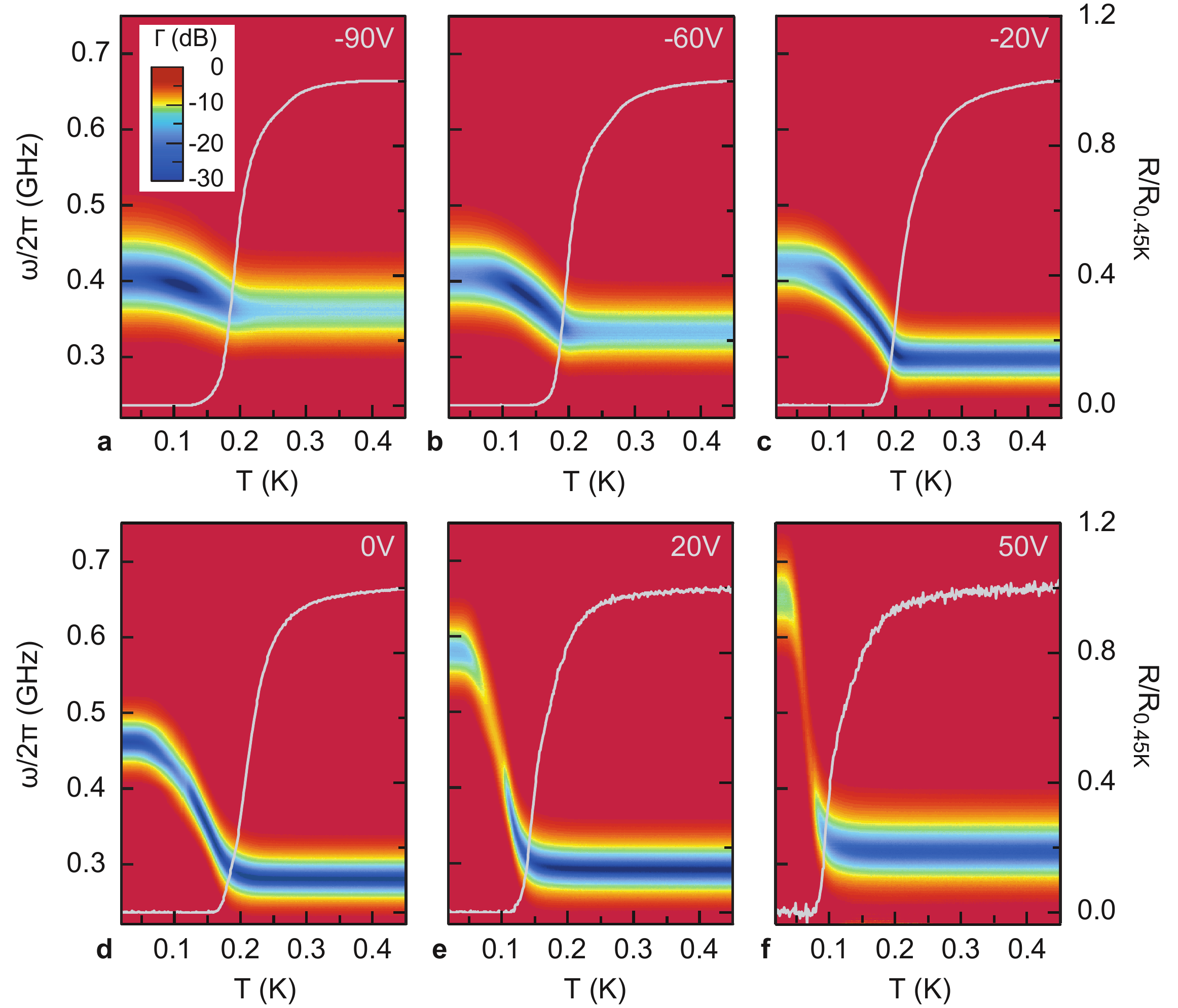}%
\end{center}
\vskip -0.5cm
\caption{\label{fig2} \textbf{Resonant microwave transport in the superconducting  state}. (a-f) Magnitude of $\Gamma(\omega)$ in dB as a function of temperature for different values of the gate voltage after calibration \cite{singh}. The normalized sheet resistance as a function of temperature is shown on the right axis.}
\end{figure*}

 We further investigated  the superconducting properties of the (110)-oriented \LAO/\STO interfaces by measuring their superfluid stiffness $J_{s}$, which characterizes the phase rigidity of the condensate. This fundamental energy scale is directly related to the imaginary part of the complex conductivity of the superconductor $\sigma _{2}(\omega)$ at finite frequency \cite{MB,dressel}
\begin{equation}\label{eq1}
J_{s}=\frac{\hbar^{2}\sigma _{2}(\omega)\omega }{4e^{2}}=\frac{\hbar^{2}}{4e^{2}L_\mathrm{k}}
\end{equation}
 where $L_\mathrm{k}$ is  the kinetic inductance of the superconductor due of the inertia of the Cooper pairs. In the normal state ($T=0.45$ K $>$ $T_c$), $\sigma_2(\omega)$=0 and the sample circuit described in Fig. 1b resonates at the frequency $\omega_{0}=\frac{1}{\sqrt{L_{1}C_\mathrm{sto}}}$,  which can be determined by measuring the reflection coefficient  of the sample circuit $\Gamma(\omega)=\frac{A^\mathrm{in}}{A^\mathrm{out}}$ (Fig. 1b). The resonance manifests itself as a dip in the magnitude of $\Gamma(\omega)$  accompanied by a $2\pi$ phase shift  (Supplementary Figure 2) \cite{singh}. Figure 2 shows the temperature dependence of the reflection coefficient for a selection of gate voltages.  In the superconducting state, the 2-DEG conductance acquires an imaginary part $\sigma_2(\omega)=\frac{1}{L_\mathrm{k}\omega}$ that generates a shift of $\omega_0$ towards high frequencies since the total inductance of the circuit becomes  $L_\mathrm{tot}(T)=\frac{L_1L_\mathrm{k}(T)}{L_1+L_\mathrm{k}(T)}$. The superfluid stiffness $J_s(T)$ is thus extracted from the resonance shift for all gate values using Eq. (1). \\
 
\begin{figure}[htpb]
\begin{center}
\vskip 0.8cm
\abovecaptionskip 5cm
\includegraphics [width=10cm]{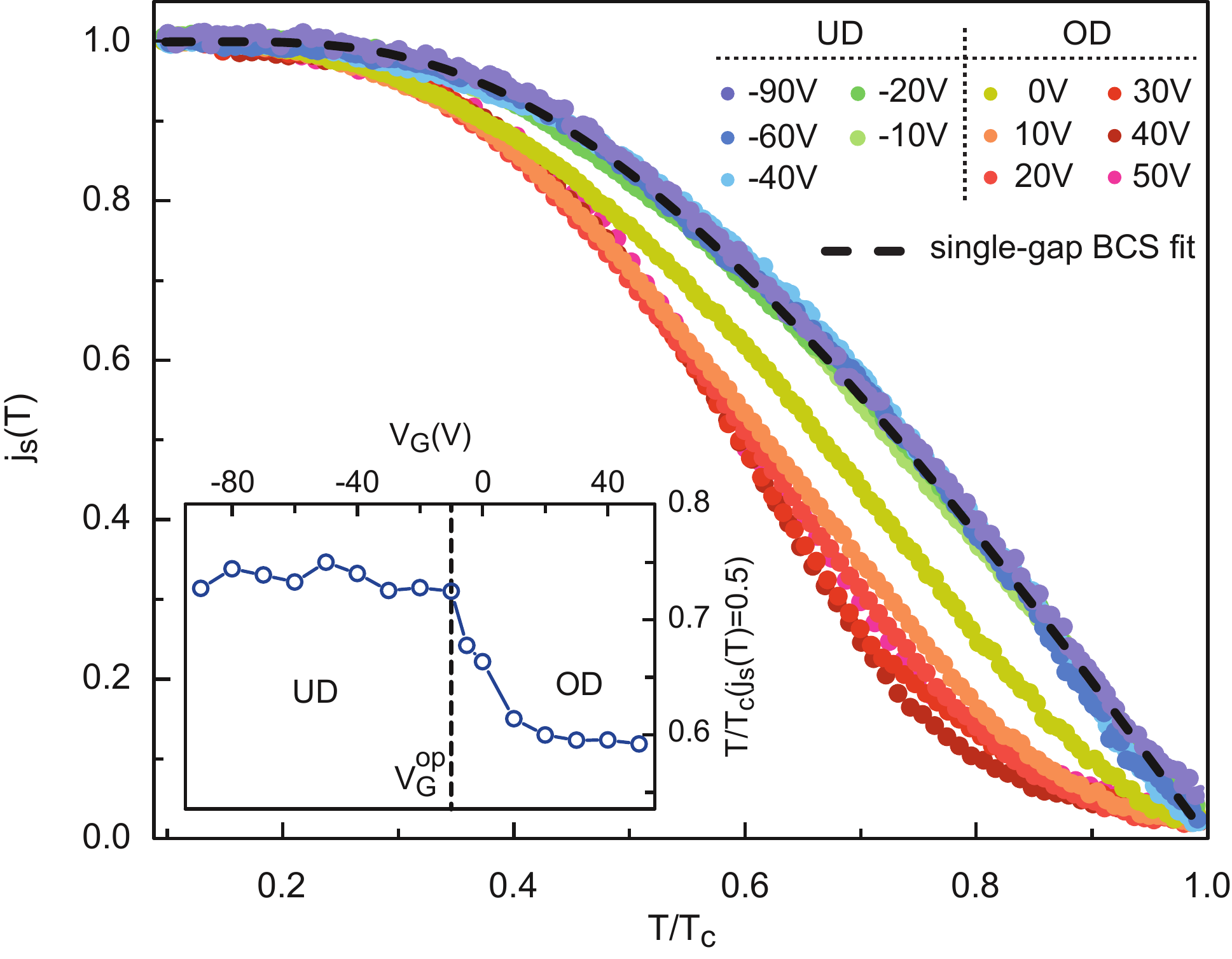}%
\end{center}
\vskip -0.5cm
\caption{\textbf{Superfluid stiffness in the underdoped and overdoped regimes}. Normalized superfluid stiffness  $j_s(T)$ as a function of the reduced temperature $\frac{T}{T_c}$ for different gate voltages spanning the underdoped (UD) and overdoped (OD) regimes. The value of $T_c$ used in the reduced temperature is the one extracted from the fit of Fig 4. In the UD regime, all the curves are superimposed and follow a single-gap BCS behavior (dashed line). In the OD regime, the temperature dependence of the $j_s(T)$ curve is strongly modified. The absolute value of $J_s$ at $T\simeq0$ as a function of $V_\mathrm{G}$ is shown in Supplementary Figure 3. Inset) Reduced temperature  corresponding to a reduction of 50$\%$ of the normalized stiffness. Whereas the values are constant in the UD regime, an abrupt decrease takes place in the OD regime.}   
\end{figure} 

 \begin{figure}[htpb]
\begin{center}
\vskip 0.8cm
\abovecaptionskip 5cm
\includegraphics [width=15cm]{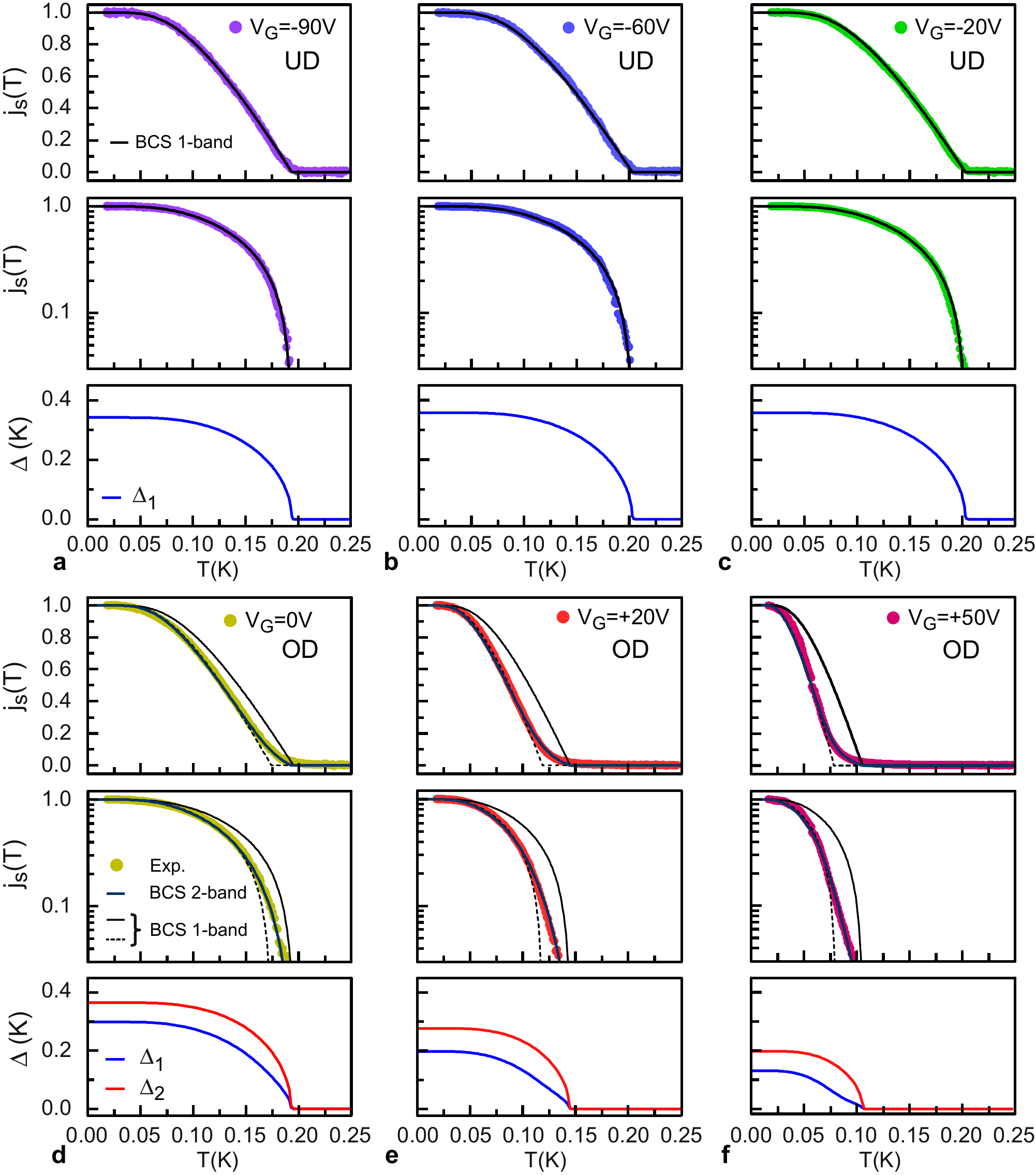}%
\end{center}
\vskip -0.5cm
\caption{\textbf{Single-band to two-band superconductivity transition in the superfluid stiffness}. Temperature dependence of $j_s$ (symbols) in linear scale (upper panels) and logarithmic scale (middle panels) for three gate voltages in the UD regime ((\textbf{a}) $V_\mathrm{G}=-90V$, (\textbf{b}) $V_\mathrm{G}=-60V$ and (\textbf{c}) $V_\mathrm{G}=-20V$) and three gate voltages in the OD regime ((\textbf{d}) $V_\mathrm{G}=0V$, (\textbf{e}) $V_\mathrm{G}=20V$ and (\textbf{f}) $V_\mathrm{G}=50V$). In the UD regime, $j_s(T)$ is fitted using the single-band model (black line) assuming in Eq. (2) the gap energy $\Delta_1(T)$ shown in the lower panel for each gate voltage. In the OD regime, $j_s(T)$ is fitted using the two-band model (blue line) corresponding to Eq. (3), assuming in Eq. (2) the gap energies  $\Delta_1(T)$ and  $\Delta_2(T)$ shown in the lower panel for each gate voltage. The values of the $\gamma$ coefficient and coupling constants are given in Supplementary Figure 5 and 6. Two attempts to fit $j_s(T)$ in the overdoped regime with different $T_c$ using a single-band model are also shown (black and dashed lines). The fit at optimal doping $V_\mathrm{G}^\mathrm{opt}=-10V$ is shown in supplementary figure 4.}
\end{figure}

For a single-band superconductor, the temperature dependence of the normalized stiffness $j_s(T)=\frac{J_s(T)}{J_s(0)}$ can be derived from the BCS superconducting gap energy $\Delta (T)$  \cite{kogan}
\begin{equation}\label{eq2}
j_s=\delta^2\sum_{n=0}^\infty[\delta^2+(n+1/2)^2]^{-3/2}
 \end{equation}
 where $\delta=\frac{\Delta (T)}{2\pi k_BT}$ is the dimensionless gap energy and $n$ is an integer that defines the Matsubara frequencies $\hbar\omega=\pi T(2n+1)$. In Figure 3 we compare the temperature dependence of $j_s$ in the underdoped (UD) and overdoped (OD) regions of the phase diagram. 
 Two distinct behaviors can be identified.  In the UD regime,  all the curves overlap each other and follow a single-gap BCS behavior (Eq. 2) in consistency with the single band Hall effect reported in Fig. 1e.  On the other hand, in the OD regime, $j_s$ shows a strong deviation with respect to the single-gap BCS behavior. In particular, a change in curvature and a prominent tail appear for $j_s<0.5$. The inset of Fig. 3 emphasizes this difference in trend by showing the reduced temperature $\frac{T}{T_c}$ corresponding to a reduction of 50$\%$ of $j_s$ as a function of gate voltage. Whereas this value is constant in the UD regime, it drops abruptly at $V_\mathrm{G}^\mathrm{opt}$ and further decreases in the OD regime. This result clearly suggests that a transition  between single-band superconductivity and two-band superconductivity occurs at $V_\mathrm{G}^\mathrm{opt}$ due to the filling of band 2 as anticipated from Hall effect (Fig. 1e).  To calculate the superfluid stiffness in  two-band superconductors, Kogan \textit{et al.} developed a self-consistent BCS approach based on the quasi-classical  Eilenberger weak coupling formalism \cite{kogan,kim2gap}. First, the two superconducting gaps $\Delta_1(T)$ and $\Delta_2(T)$ are  self-consistently calculated by introducing the intraband ($\lambda_{11}$, $\lambda_{22}$) and interband ($\lambda_{12}$, $\lambda_{21}$) coupling constants (Supplementary Note 1).   Then, the temperature dependence of the normalized stiffness for each band $j_{s1}$ and $j_{s2}$ is obtained by introducing the corresponding gaps $\Delta_1$ and $\Delta_2$ in Eq (2). The total normalized stiffness is the sum of the contributions of each band \cite{kogan}
\begin{equation}
j_s=\gamma j_{s1}+(1-\gamma)j_{s2}
 \end{equation}
where the $\gamma$ coefficient accounts for the weight of each band in the superfluid condensate. A systematic fitting procedure based on this approach was applied to $j_s$. As shown in Fig. 4 for different gate voltages spanning both the UD regime (panels a,b and c) and the OD regime (panels d,e and f), a very good agreement is obtained between the experimental data and the BCS model. In particular, the transition between single-band and two-band superconductivity is clearly visible both in linear scale (up sub-panels) and logarithmic scale (middle sub-panels). In the OD regime, the change in curvature and the tail in $j_s{(T)}$ curves are well described by the model considering a weak interband coupling ($|\lambda_{12(21)}|<<\lambda_{11(22)}$). In this case, the largest gap ($\Delta_2$) is only weakly affected by this coupling and behaves essentially as a single BCS gap closing at $T_c$ (low sub-panels). On the other hand, the smallest gap ($\Delta_1$)  follows a BCS trend at low temperature but instead of closing at a correspondingly lower transition temperature, it extends to $T_c$. This  generates a change in curvature and a tail which is further reproduced in $j_s(T)$. The fits implies that the band with the smallest gap and the tail contributes to the total superfluid stiffness with the strongest weight.  According to the band hierarchy in the quantum well, this must correspond to band 1, which is mainly filled. Fig. 5c summarizes the values of the superconducting gap energies in the phase diagram extracted from the fitting procedure. In the underdoped regime, $\Delta_1(0)$ very weakly increases with $V_\mathrm{G}$, which is consistent with BCS theory since for a 2D superconductor, $T_c$ is  in principle carrier density independent (in absence of other mechanisms affecting $\lambda_{11}$). In the overdoped regime, band 2 is also filled, and a second superconducting gap $\Delta_2$ opens at the Fermi surface, which is larger than $\Delta_1$.  The correspondence of these bands to specific orbital symmetries is discussed below. \\

\begin{figure}[t]
\begin{center}
\vskip 0.8cm
\abovecaptionskip 5cm
\includegraphics [width=10cm]{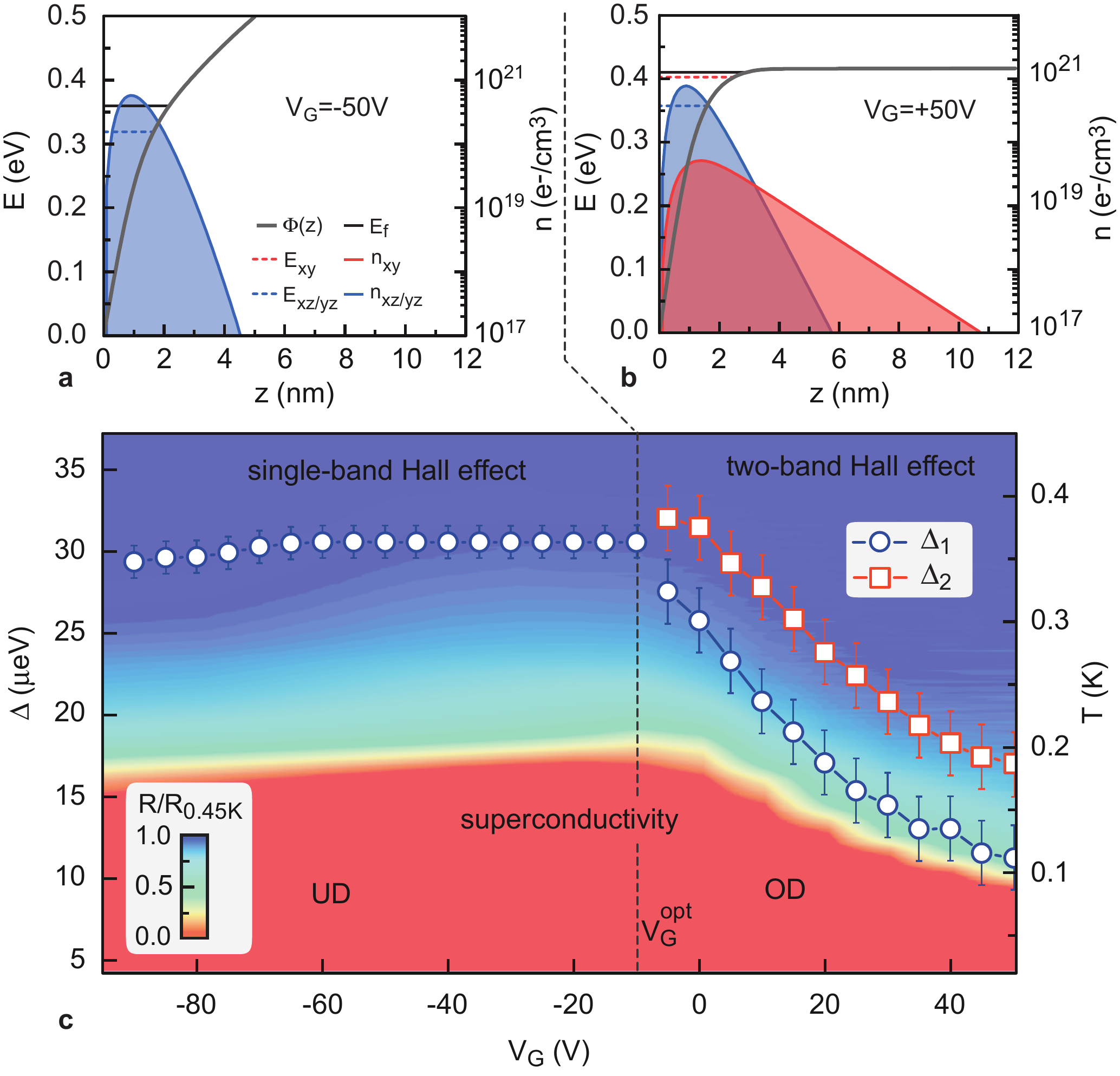}%
\end{center}
\vskip -0.5cm
\caption{\textbf{Superconducting phase diagram}. a,b) Numerical simulations of the band structure in the (110)-oriented interfaces which shows the energies of the different $t_{2g}$ bands in the confinement potential $\Phi(z)$ (left axis)  and the 3D carrier densities ($n_{xy}$,$n_{xz/yz}$) associated to each band (right axis), obtained by solving the self-consistent Poisson-Schr\"odinger equations for two doping regime : $V_\mathrm{G}$=-50V (panel a) and  $V_\mathrm{G}$=+50V (panel b). The 2D carrier density used in the simulation corresponds to the one reported  on Fig. 1e, i. e. $n\simeq$0.75 10$^{14}$e/cm$^2$ at $V_\mathrm{G}$=-50V and $n\simeq$1.1 10$^{14}$e/cm$^2$ at $V_\mathrm{G}$=+50V. See Supplementary Note 2 for details on the numerical simulation. c) Gap energies $\Delta_1(0)$ and $\Delta_2(0)$ extracted from the fitting procedure (left axis) as a function of $V_\mathrm{G}$, superimposed on the sheet resistance (color scale) of the 2-DEG plotted as a function of temperature (rigth axis) and $V_\mathrm{G}$.}
\end{figure}

Arguing that superconductivity in \STO-based interfaces has likely the same origin than that of bulk \STO, early reports of a distinctive double-gap structure in the tunneling density of states in Nb-doped bulk \STO above a certain density ($\simeq$ 5 $\times$ 10$^{19}$ cm$^{-3}$) \cite{binnig} motivated the theoretical suggestion that 
multiband superconductivity can also take place also in  \LAO/\STO heterostructures \cite{trevisan,fernandes}.  While the onset of multiple-band occupancy in bulk \STO has been recently confirmed by the analysis of quantum oscillations in the normal state \cite{benhia14}, very recent tunneling \cite{swartz} and a.c. conductivity experiments do not show any evidence for multi-gap supercondutivity \cite{scheffler}. For heterostructures, only the conventional (001)-orientation has been theoretically considered, and in this type of interfaces, experiments \cite{richter,singh,bert2} are rather consistent with the presence of a single superconducting gap. To establish a relation between the two gaps evidenced in our  experiment and the different $t_{2g}$ bands, we numerically simulated the interfacial band structure by self-consistently solving  the Poisson-Schr\"odinger equations (Fig. 5a-b)(see Supplementary Note 2). In this crystal  orientation, the band hierarchy is reversed with respect to the conventional (001)-orientation \cite{pesquera,gervasi}. The degenerated $d_{xz}$/$d_{yz}$ bands are always filled whereas the $d_{xy}$ band at higher energy is only occupied at strong electrostatic doping. The ability of these different bands to host superconductivity critically depends  on their density of states $N(0)$, which is directly proportional to their in-plane effective mass. In contrast to the (001)-oriented interfaces, where the low density of states of the $d_{xy}$ band precludes the formation of superconductivity at low doping, in the (110)-orientation, the two bands have relatively high and similar DOS ($m^{xz,yz}_{\parallel}\simeq2.3m_0$ and $m^{xy}_{\parallel}\simeq3.1m_0$, see Supplementary Note 2). Consequently, they are both suitable to sustain superconductivity, which explains why superconductivity cannot be (or is hardly) suppressed by carrier depletion, and why two-band superconductivity is observed at high electrostatic gating. According to the numerical simulations, we relate band 1 ($\Delta_1$, $J_{s1}$) to the $d_{xz}$/$d_{yz}$ band and band 2 ($\Delta_2$, $J_{s2}$) to  the $d_{xy}$ band.   \\

One of the fundamental questions that arises in two-band superconductors is whether the energy gaps in the different bands have the same sign.  The temperature dependence of the superfluid density suggests that there are no nodes, so the gap must be constant over each of the two Fermi surfaces. On the other hand, the two superconducting order parameters can have the same sign ($s_{++}$-wave pairing) for an attractive interband interaction ($\lambda_{12(21)}>0$) or opposite sign  ($s_\pm$-wave pairing) for a repulsive interband interaction ($\lambda_{12(21)}<0$).  This issue has been widely discussed in the context of iron-based superconductors, where antiferromagnetic spin fluctuations are expected to provide a repulsive  interband pairing channel \cite{wang,mazin,hirschfeld}, leading to $s_\pm$ pairing with two nodeless gaps of opposite signs on the hole and electron pockets.  The presence of multiple gaps in these systems has been confirmed by several thermodynamic probes, like e.g. the temperature dependence of the superfluid density \cite{prozorov-review}. Recently, the analysis of quasiparticle-interference imaging in STM experiments also allowed to demonstrate the sign-changing nature of the order parameter, confirming the $s_\pm$ nature of superconductivity in Fe(Se,Te) and FeSe \cite{hanaguri,sprau}.  According to BCS theory, only the positive quantity $\lambda_{12(21)}^2$ affects the determination of $T_c$. Therefore, the coupling with an additional band with a significant density of states should always lead to an increase of $T_c$, regardless of the sign of $\lambda_{12}$ \cite{kogan,fernandes}.  This would then be in contradiction with our findings, where $T_c$ decreases soon after that the second band is populated, see Fig. 5. However, this apparent discrepancy could be reconciled by accounting for impurity scattering in a $s_\pm$ superconductor. Indeed, although nodeless single-band superconductivity is essentially insensitive to scattering (provided it is not too strong),  in multiband $s_\pm$ superconductors scattering processes between bands with opposite-sign gaps are pair-breaking, leading to a suppression of superconductivity \cite{kogan2,fernandes}. The mechanism is similar to the one occurring in high-$T_c$ superconducting cuprates,  where the  $d_{x^2-y^2}$ superconducting order parameter changes sign between different regions of the single-band Fermi surface, making impurity scattering detrimental for superconductivity. In this view the $s_\pm$ symmetry of the order parameter in our (110)-oriented \LAO/\STO interfaces could explain why the onset of superconductivity in the second band makes the system more sensitive to impurity scattering, leading to the unusual suppression of $T_c$ observed in the OD regime. It is worth noting that the analysis of the temperature dependence of the superfluid density based on Eq.\ (\ref{eq2}) relies mainly on the relative band filling and on the gap values, so it is a robust finding rather insensitive to the presence of disorder. On the other hand, the suppression of the gaps and $T_c$ in the OD regime, that we simply modeled via a decrease of the  superconducting couplings $\lambda_{ij}$ in the absence of disorder, could be understood as the effect of disorder within a multiband model with constant pairing interactions and $s_\pm$ gap symmetry\cite{trevisan}.  Although the origin of the repulsive interband coupling in our system remains to be identified, we anticipate that it could be related to the specific $t_{2g}$ orbitals symmetry of the bands.

	 \indent According to the typical gap energies reported in Fig. 5c, the transition from single-band to  two-band superconductivity in the (110)-oriented \LAO/\STO interface, can be confirmed by direct tunnelling spectroscopy  using planar junctions  or  STM spectroscopy. However, the relative sign must be investigated using of a phase-sensitive experiment  such as a Josephson experiment for example \cite{chen}. Finally, we also mention that a similar situation can occur in the conventional  (001)-oriented interface whose superconducting phase diagram $T_c(V_\mathrm{G})$ displays a characteristic dome-shape \cite{Caviglia:2008p116,biscaras2,Bell:2009p6086}. Although tunnelling spectroscopy mainly reveals a single gap \cite{richter}, we cannot exclude that the decrease of $T_c$ in the overdoped regime could originate from a repulsive coupling with a reversed-sign superconducting gap that opens in a high-energy $d_{xy}$ replica band \cite{trevisan}. Because $N(0)$ is weak for the $d_{xy}$ bands in this orientation, interband scattering can contribute to the reduction of $T_c$  but  the gap may not be easily visible in either tunnelling spectra nor in superfluid stiffness. \\

 
\textbf{Acknowledgments}\\
We acknowledge K. Behnia for useful discussions. This work has been supported by the R\'egion Ile-de-France in the framework of CNano IdF, OXYMORE and Sesame programs, by CNRS through a PICS program (S2S) and ANR JCJC (Nano-SO2DEG). This work was supported by the Spanish MAT2017-85232-R, MAT2014-56063-C2-1-R,  Severo Ochoa SEV-2015-0496 grant, and the Generalitat de Catalunya (2017 SGR 1377). This work has been supported by the Italian MAECI under the Italia-India collaborative project SUPERTOP-PGR04879. \\

\indent

\newpage
\large\textbf{Methods}\\

\textbf{Sample preparation}\\
 The 10 u.c. thick \LAO film was grown by pulsed laser deposition ($\lambda$= 248 nm) monitored by reflection high-energy electron diffraction (RHEED). The substrate was heated from room temperature to deposition temperature (850$^\circ$C) in an oxygen partial pressure P(O$_2$) = 0.1 mbar. During deposition, the \LAO was grown under a pressure P(O$_2$) = 10$^{-4}$ mbar and a 1 Hz repetition rate, with laser pulse energy of around 26 mJ. At the end of the deposition, the sample was cooled down in oxygen rich atmosphere to minimize the formation of oxygen vacancies that could lead to extrinsic mechanisms of conduction. More specifically, the samples were cooled from T = 850$^\circ$C to 750$^\circ$C under a pressure P(O$_2$)= 0.3 mbar and under P(O$_2$) = 200 mbar from T = 750$^\circ$C down to room temperature, including a dwell time of 1 hour at 600$^\circ$C.\\

\textbf{Hall effect and gate capacitance. }\\
The (110)-oriented \LAO/\STO interface displays a Hall effect linear with magnetic field for V$_\mathrm{G}<V_\mathrm{G}^\mathrm{opt}$ =-10V as expected for single band transport.  For V$_\mathrm{G}>V_\mathrm{G}^\mathrm{opt}$, a nonlinear Hall effect is observed due to the filling of a second band (Supplementary Figure 1) \cite{singhCR,biscaras2}. The carrier density $n_\mathrm{Hall}=\frac{B}{eR_\mathrm{Hall}}$, reported in Fig.1e of the main text, which is extracted from $R_\mathrm{Hall}$ in the limit B$\rightarrow$0, is only meaningful in the linear regime. However, the correct variation of carrier density as a function of $V_\mathrm{G}$ can be retrieved from the charging curve of the capacitor given by the integral of the gate capacitance $C(V_\mathrm{G})$ :
\begin{equation}
n_\mathrm{2D}(V_\mathrm{G})=n_\mathrm{Hall}(V_\mathrm{G}=-120V)+\frac{1}{eA}\int_{-120}^{V_\mathrm{G}}C(V_\mathrm{G})dV
\end{equation}
where $A$ is the area of the sample. As shown in Figure 1 of the main text, $n_\mathrm{2D}$ matches  $n_\mathrm{Hall}$ in the single-band regime ($V_\mathrm{G}<V_\mathrm{G}^\mathrm{opt}$) and extrapolates the curve in the two-band regime.\\

\textbf{Reflection coefficient measurement and calibration}\\
A directional coupler is used to guide the microwave signal from the input port to the sample through a bias-tee, and to separate the reflected signal which is amplified by a low-noise cryogenic HEMT amplifier before reaching the output port \cite{singh}.  The complex transmission coefficient $S_{21}(\omega)$ between the two ports is measured with a vector network analyzer.   Standard microwave network analysis relates the reflection coefficient of the RLC sample circuit $\Gamma(\omega)$ to the measured $S_{21}(\omega)$, through complex error coefficients that can be determined by a calibration. In this experiment, the microwave set-up was calibrated by using as references, the impedances $Z_\mathrm{c}$ of the sample circuit in the normal state of the 2-DEG for different gate values. 
 The complex reflection coefficient $\Gamma(\omega)$ is given by 
\begin{eqnarray}
\Gamma(\omega)=\frac{A^\mathrm{out}(\omega)}{A^\mathrm{in}(\omega)}=\frac{Z_c(\omega)-Z_0}{Z_c(\omega)+Z_0} 
\label{reflect}
\end{eqnarray}
 where   $A^\mathrm{in}$ and  $A^\mathrm{out}$ are the complex amplitudes of incident and reflected waves and $Z_0$ = 50 $\Omega$ is the characteristic impedance of the CPW transmission line. A reflection measurement gives therefore a direct access to the circuit impedance $Z_c(\omega)$. Supplementary Figure 2 shows the magnitude of the reflection coefficient $\Gamma$ as a function of frequency and gate voltage at T=450 mK. The resonance manifests itself as a dip in the magnitude of $\Gamma(\omega)$  accompanied by a $2\pi$ phase shift. The value of $C_{sto}$ which is gate dependent because of the electric field dependence of the dielectric constant of the \STO substrate can be determined from the resonance frequency $\omega_0=\frac{1}{\sqrt{L_1C_{sto}}}$for each value of the gate voltage. \\
 
\newpage
\begin{center}
\textbf{Supplementary Material}\\
\end{center}
 \textbf{Supplementary Note 1 : Two-band superfluid stiffness analysis}\\
 
The temperature dependence of the normalised superfluid stiffness in the OD regime was analysed using the model developed in reference \cite{kogan}. We consider two superconducting bands with density of states $N_{1,2}$ and interband coupling constants $\lambda_{11(22)}$. In addition, we add a small interband coupling between the two bands characterised by the constants $\lambda_{12(21)}$. In this BCS approach, the superconducting $T_c$ is determined by the following equation :
\begin{equation}
1.76k_BT_c=2E_D\exp(-\frac{1}{\tilde{\lambda}})
\end{equation}
where $E_D$=400K is the Debye energy \cite{maniv} and $\tilde{\lambda}$ the effective coupling constant given by
\begin{equation}
\tilde{\lambda}=\frac{2(\lambda_{11}\lambda_{22}-\lambda_{12}\lambda_{21})}{\lambda_{11}+\lambda_{22}-\sqrt{(\lambda_{11}-\lambda_{22})^2+4\lambda_{12}\lambda_{21}}}
\end{equation}
For the interband coupling terms we assume that $\frac{\lambda_{12}}{n_2}=\frac{\lambda_{21}}{n_1}$, where $n_{1}=\frac{N_{1}}{N_1+N_2}$ $\simeq$ 0.6 and $n_{2}=\frac{N_{2}}{N_1+N_2}$ $\simeq$ 0.4  are the density of states weight in each band, which are determined from the average in-plane masses (see Supplementary Note 2).
 We introduce the dimensionless gap energies
\begin{equation}
\delta_{\nu}=\frac{\Delta_\nu}{2\pi T}
 \end{equation}
which is given by solving self-consistently the two following equations 
\begin{equation}
\delta_{\nu}=\sum_{\mu=1,2}\lambda_{\nu\mu}\delta_\mu\Big(\frac{1}{\tilde{\lambda}}+\ln\frac{T_c}{T}-A_\mu\Big)
 \end{equation}
\begin{equation}
A_\mu=\sum_{n=0}^{\infty}\Big(\frac{1}{n+1/2}-\frac{1}{\sqrt{\delta_\mu^2+(n+1/2)^2}}\Big)
 \end{equation}
 For each band the temperature dependence of the associated normalized stiffness is derived from the dimensionless gap energy $\delta_{\nu}$ 
 \begin{equation}\label{eq2}
j_{s\nu}=\delta_{\nu}^2\sum_{n=0}^\infty[\delta_{\nu}^2+(n+1/2)^2]^{-3/2}
 \end{equation}
 The total normalized stiffness is the sum of the contributions of each band 
\begin{equation}
j_s=\gamma j_{s1}+(1-\gamma)j_{s2}
 \end{equation}
where the $\gamma$ coefficient accounts for the weight of each band in the superfluid condensate. Data shown in Figure 4 of the main text has been fitted using the relations (6) and (7). Note that in this pure BCS approach, the effect of fluctuations has not been included. Supplementary Figure 5 shows the evolution of the $\gamma$ coefficient with the gate voltage. \\
 
 In Supplementary Figure 6 we show the doping dependence of the dimensionless coupling constant extracted from the analysis of the superfluid stiffness. As explained in the method section, $\tilde\lambda$ is the effective coupling determining the $T_c$. In the single-band case, it coincides with $\lambda_{11}$, while for the multiband regime it is given by Eq. (2). As mentioned in the main text, the analysis of the temperature dependence of the superfluid density relies mainly on the relative band filling and on the gap values, so it is a robust finding rather insensitive to the presence of disorder. On the other hand, $\Delta_\nu(T)$ must be derived using a specific model for the superconducting state. To simplify the analysis, we relied on the Eq.s (3)-(5), where the gap are computed for a multiband model in the absence of disorder. In this case, the only way to obtain a decrease of the gap values at $T=0$ is to reduce the coupling constants for increasing density, as shown in In Supplementary Figure 6.  However, a similar result could be possibly obtained by using constant superconducting couplings and introducing the effect of disorder. Indeed, as recently discussed in Ref. \cite{trevisan}, in the presence of disorder a $s_\pm$ superconducting state is strongly affected by the onset of occupancy of the secondary band. Indeed, interband impurity scattering is pair breaking in a $s_\pm$ superconductor, leading to a gradual suppression of both the critical temperature and of the superconducting gap across the Lifshitz transition. A quantitative analysis of this effect requires however a detailed analysis of a dirty multiband model that is well beyond the scope of the present work. \\
 
 \textbf{Supplementary Note 2 : Interfacial band structure simulation }\\
 
The filling and the confinement of each $t_{2g}$ band are determined by solving self-consistently the schr\"odinger and the Poisson equations as described in the supplementary material of reference \cite{biscaras2}.  The hopping terms in Fig 1a. were taken to be $t_1$=-320meV and $t_2$=-16meV,  which correspond to the masses $m^{xz,yz}_{001}=-\frac{\hbar^2}{2a^2t_1}$ $\simeq$ 0.7m$_0$ for the $d_{xz/yz}$ orbitals and $m^{xy}_{001}=-\frac{\hbar^2}{2a^2t_2}$ $\simeq$ 14m$_0$ for the $d_{xy}$ orbitals along the (001) direction of the conventional (001)-oriented interfaces. In the (110)-oriented interfaces, the masses along the confinement direction (i. e. the (110) direction)  are $m^{xz,yz}_{110}=-\frac{\hbar^2}{2a^2(t_1+t_2)}$ $\simeq$ 0.65m$_0$  and $m^{xy}_{110}=-\frac{\hbar^2}{4a^2t_1}$ $\simeq$ 0.35m$_0$ \cite{wang110}. The density of states can be determined from the in-plane masses in the (1-10) and (001) directions. For the $d_{xy}$ orbital, $m^{xy}_{001}=-\frac{\hbar^2}{2a^2t_2}$ $\simeq$ 14m$_0$ and $m^{xy}_{1-10}=-\frac{\hbar^2}{2a^2t_1}$ $\simeq$ 0.7m$_0$, corresponding to an average in-plane mass  $m^{xy}_{\parallel}=\sqrt{m^{xy}_{001}m^{xy}_{1-10}}$ $\simeq$ 3.1m$_0$ \cite{wang110}.
For the $d_{xz,yz}$ orbitals, $m^{xz,yz}_{001}=-\frac{\hbar^2}{2a^2t_1}$ $\simeq$ 0.7m$_0$ and $m^{xz,yz}_{1-10}=-\frac{\hbar^2}{2a^2\big(\frac{2t_1t_2}{t_1+t_2}\big)}$  $\simeq$ 7.3m$_0$, corresponding to an average in-plane mass  $m^{xz,yz}_{\parallel}=\sqrt{m^{xz,yz}_{001}m^{xz,yz}_{1-10}}$ $\simeq$ 2.3m$_0$ \cite{wang110}. \\

\begin{figure}[h!]
\includegraphics[width=8cm]{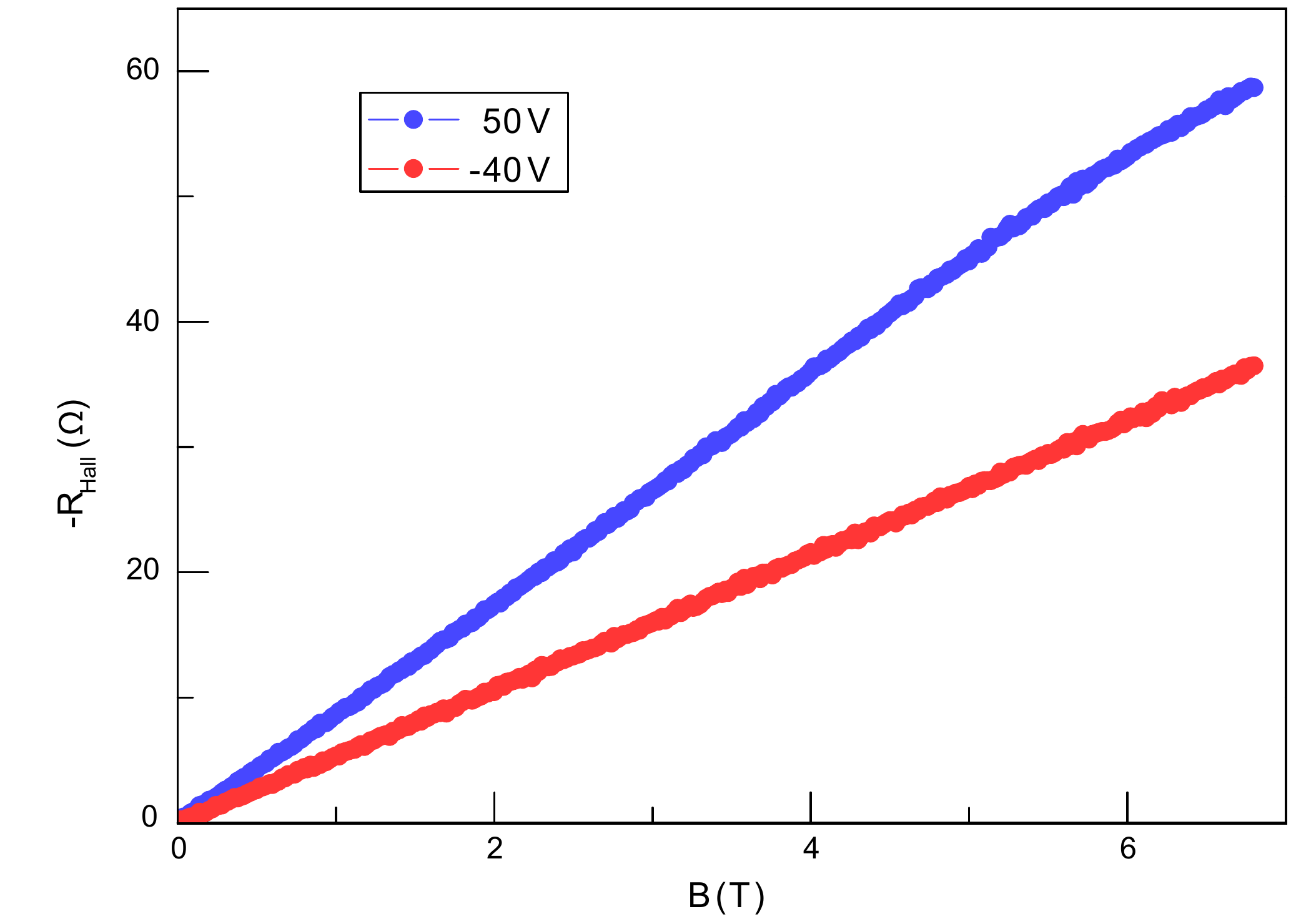}
\caption{Hall resistance as a function of magnetic field applied perpendicular to the interface for $V_\mathrm{G}$ = -40V (single-band) and for $V_\mathrm{G}$ = 50V (two-band).}
\label{FigureS1}
\end{figure}
\begin{figure}[h!]
\includegraphics[width=8cm]{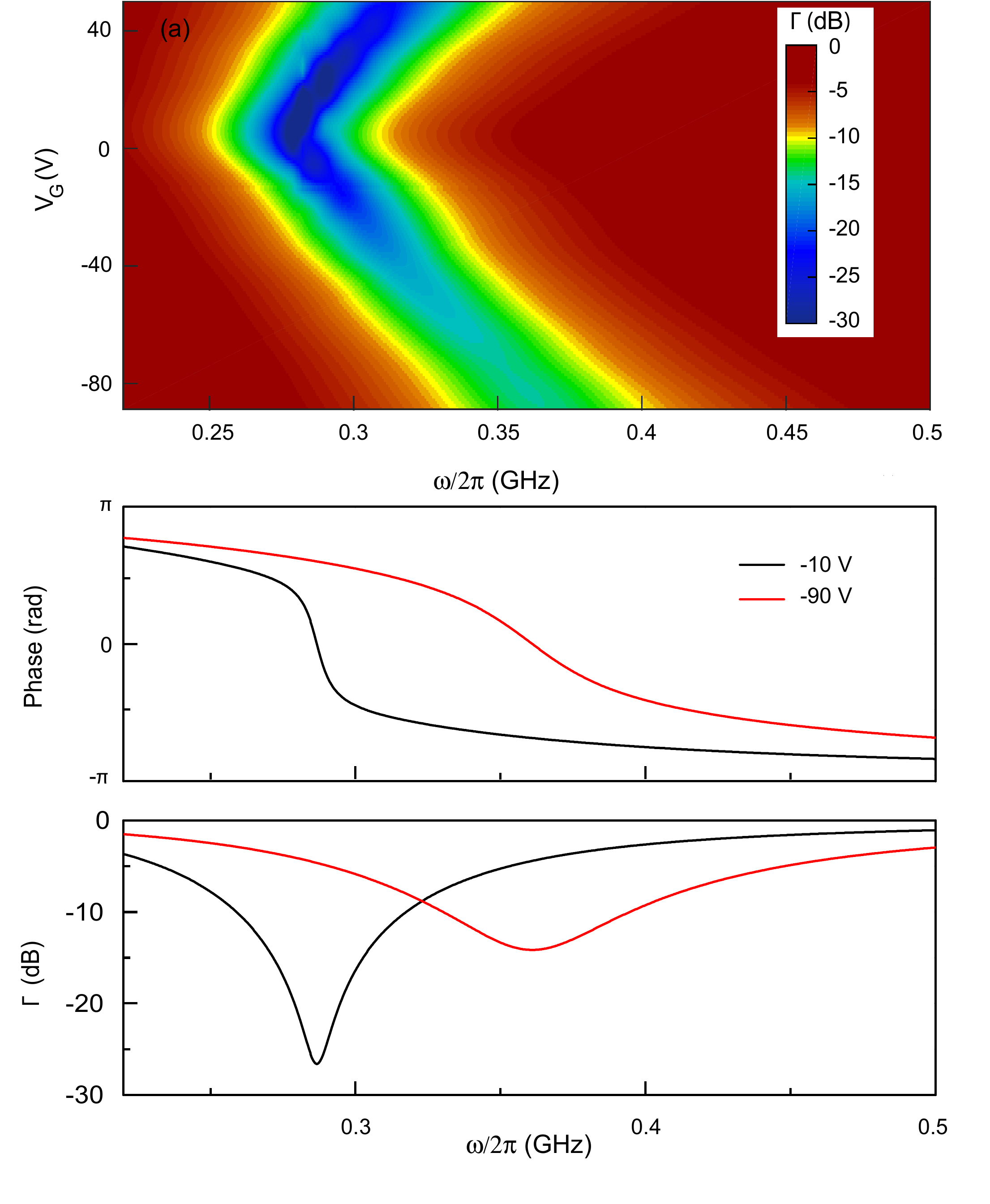}
\caption{(a) Magnitude of $\Gamma$ in dB (color scale) as a function of $\omega$ and $V_\mathrm{G}$. $C_{sto}$ takes a maximum value of 36 pF for $V_\mathrm{G}$ = 0 V. Phase (b) and magnitude (c) of $\Gamma$ for  $V_\mathrm{G}$ =-90V and $V_\mathrm{G}$=-10V.}
\label{FigureS2}
\end{figure}
\begin{figure}[h!]
\includegraphics[width=8cm]{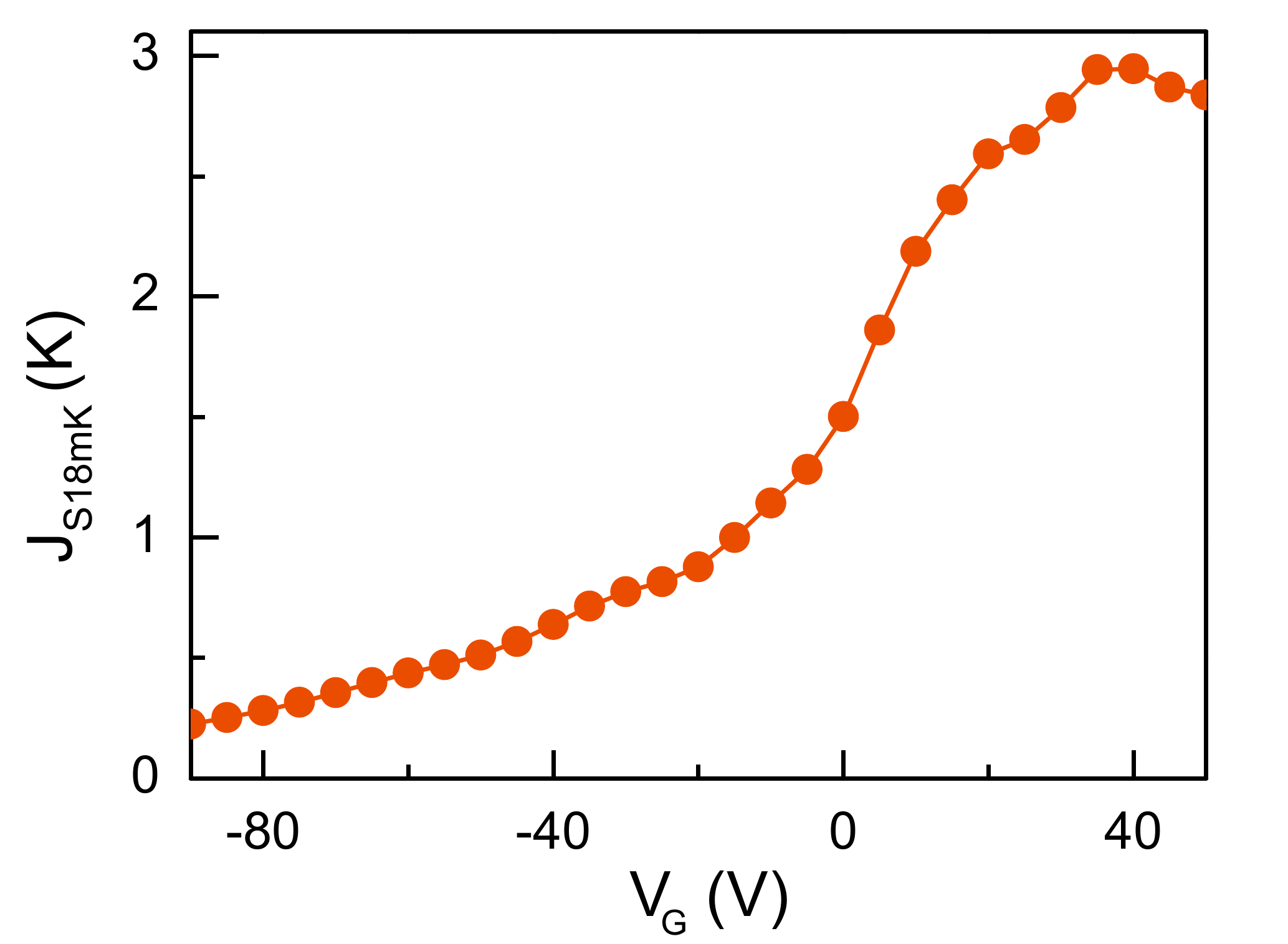}
\caption{Gate dependence of  $J_s$ extracted from $\Gamma (\omega)$ at the lowest temperature T$\simeq$18mK.  }
\label{FigureS3}
\end{figure}

\begin{figure}[h!]
\includegraphics[width=8cm]{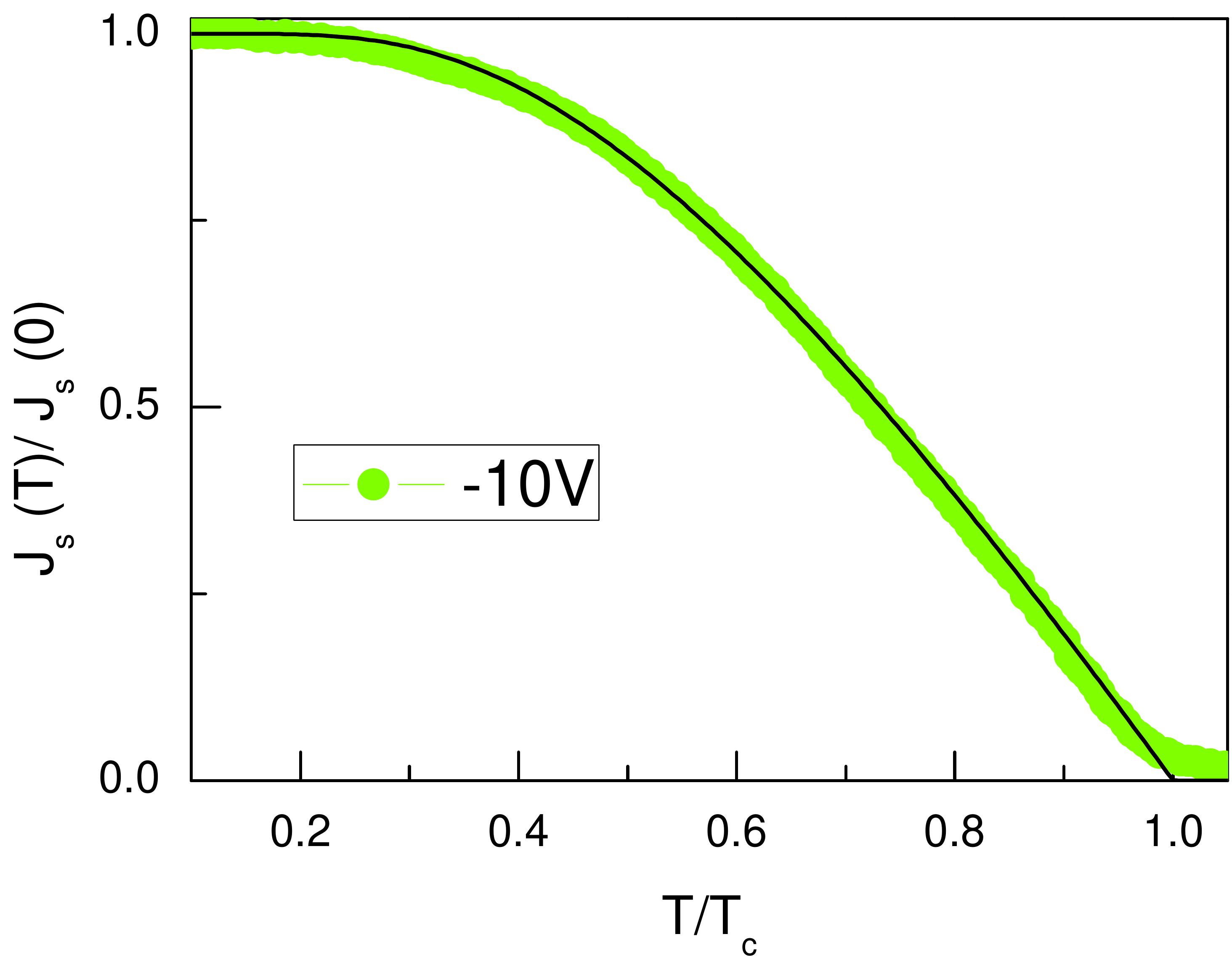}
\caption{Temperature dependence of normalised superfluid stiffness $j_s(T)$ at the optimal doping point $V_\mathrm{G}$=-10V fitted by a single band BCS model.}
\label{FigureS4}
\end{figure}

\begin{figure}[h!]
\includegraphics[width=8cm]{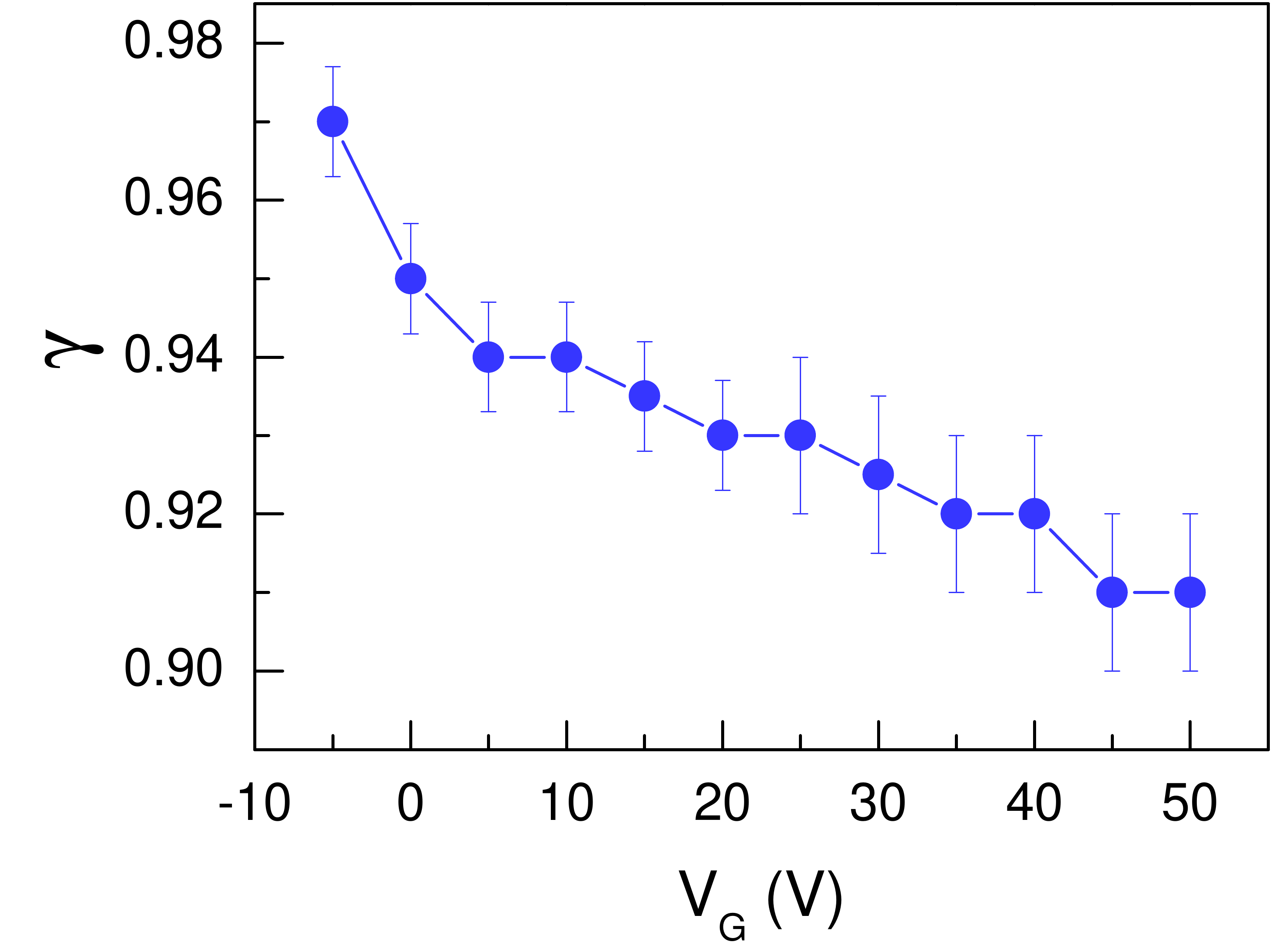}
\caption{Gate dependence of the $\gamma$ coefficient controlling the weight of each band in the superfluid stiffness (Eq. (3) in the main text).}
\label{FigureS5}
\end{figure}

\begin{figure}[h!]
\includegraphics[width=10cm]{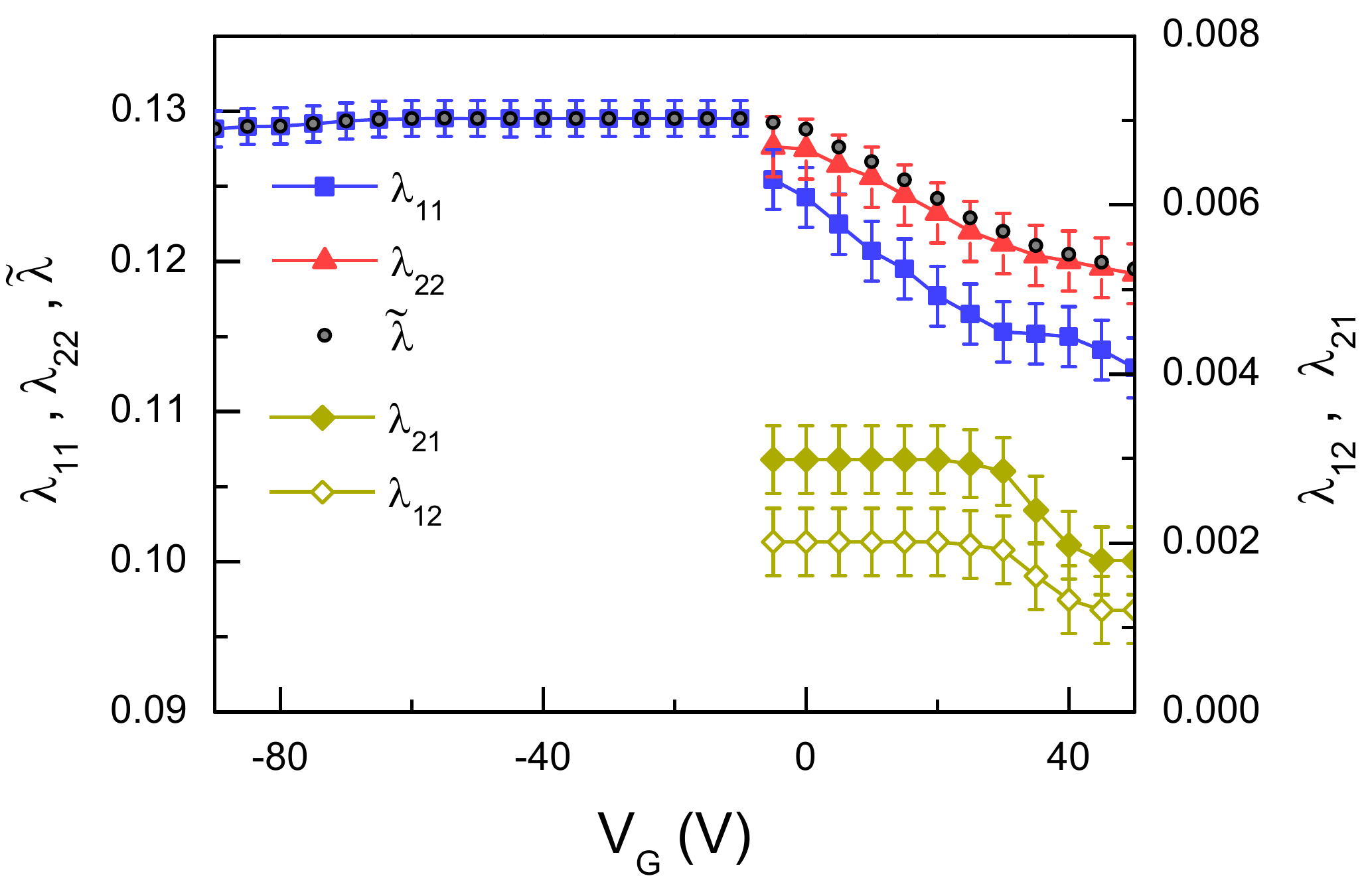}
\caption{Gate dependence of the coupling constants in the phase diagram. The main source of pairing is the positive intraband coupling $\lambda_{11(22)}$ with a small interband coupling $\lambda_{12(21)}$ which can be either positive (attractive) or negative (repulsive). The values of the coupling constants  are found to be comparable to the rare values reported in the literature for bulk \STO \cite{fernandes}}.
\label{FigureS6}
\end{figure}

\end{document}